\documentclass[%
 reprint,
 amsmath,amssymb,
 aps,
pra,
]{revtex4-2}

\usepackage{graphicx}
\usepackage{dcolumn}
\usepackage{bm}

\begin{document}

\preprint{APS/123-QED}

\title{Deterministic Preparation of Non-Gaussian Quantum States: Applications in Quantum Information Protocols}%

\author{Anindya Banerji$^{1}$}
 \affiliation{1-Centre for Quantum Technologies, National University of Singapore, Singapore.
}

\author{Graciana Puentes$^{2,3}$}%
\affiliation{2-Departamento de Fsica, Facultad de Ciencias Exactas y Naturales,
Universidad de Buenos Aires, Ciudad Universitaria, 1428 Buenos Aires, Argentina\\
3-CONICET-Universidad de Buenos Aires, Instituto de Fsica de Buenos Aires (IFIBA), Ciudad Universitaria,
Buenos Aires, Argentina.
}%

\date{\today}

\begin{abstract}
We report a scheme for deterministic preparation of non-Gaussian quantum states on-demand. In contrast to   probabilistic approaches for preparation of non-Gaussian quantum states, conditioned on photon subtraction or addition, we present a scheme that can  prepare non-Gaussian quantum states on-demand, by applying a unitary transformation which removes the Gaussianity of measurement statistics of field quadratures, namely a quadrature rotation via transmission through a beam-splitter, using a two-mode photon-number squeezed state as input. The resulting state exhibits a quantum vortex structure in quadrature space, confirming its non-Gaussian nature. Such non-Gaussian quantum state also reveals increased entanglement content, as quantified by the Logarithmic Negativity and the Wigner function negative volume, therefore displaying high potential for applications in quantum information protocols, in particular for applications in entanglement distillation schemes. 
\end{abstract}

\maketitle


\section{Introduction}

Gaussian quantum states are defined as the states for which measurement statistics of field quadratures are Gaussian. As such, they can be fully described by their mean field and covariance matrix. In the complete space of states of Continuous Variable (CV) systems, Gaussian quantum states play a key role, namely of all quantum states with a given  covariance matrix,  Gaussian states have the least entanglement  and the highest entropy \cite{12,13,3}. From a theoretical stand, Gaussian quantum states provide for a standard framework for quantum information theory. On an experimental level, CV quantum information has long been promoted due to the capability of on-demand generation of large entangled states using either time-frequency modes \cite{17,18,19,20,21,22} or spatial modes \cite{14,15,16}.\\

Despite the experimental and theoretical advantages of Gaussian states, they present  major limitations in the context of quantum technologies: all Gaussian measurements of such states can be efficiently simulated \cite{23,24,25}. Pioneer works on CV quantum computing argue that a non-Gaussian operation, meaning an operation removing the Gaussian statistics of the states, is required for implementation of a universal CV quantum computer \cite{26}. Later works laying the groundwork for CV quantum computation have left the question of such non-Gaussian operation somewhat open \cite{27,28,29}. Common schemes based on cubic phase gates, turn out to be particularly hard to implement in an experimental setup \cite{30}. Furthermore, these protocols require highly non-Gaussian states to encode information \cite{31}. In spite the fact that such quantum states could serve as a non-Gaussian resource for implementation of non-Gaussian gates \cite{32}, they remain notoriously challenging to prepare. Despite the practical problems involved with the non-Gaussian regime, non-Gaussian states are expected to provide for enhanced entanglement content, and eventually it is expected to be necessary to venture into non-Gaussian territory to reach a quantum computational advantage, in the CV regime \cite{33}.\\

Experimentally, the preparation of a set of modes in  a non-Gaussian quantum state is generally much harder than the preparation of their Gaussian counterparts. In essence, it suffices to apply a non-Gaussian unitary operation to create a quantum state with non-Gaussian statistics of field quadratures. In practice, such non-Gaussian unitary transformations are hard to come by, meaning that often  different preparation techniques are required. There are two main approaches to reach non-Gaussianity. The first approaches are probabilistic, and rely in performing non-Gaussian measurements on a Gaussian state conditioned to a certain measurement outcome, such is the case of photon-subtraction or photon-addition techniques. The second approach concentrates on deterministic methods, which rely on the implementation of non-Gaussian unitary transformations on-demand, meaning a unitary transformation that removes the Gaussianity of the quantum states in a deterministic fashion. \\

In this article, we introduce a deterministic method for preparation non-Gaussian quantum  states on demand, by implementation of a unitary transformation, namely transmission through a  beam-splitter, on an initial two-mode photon-number squeezed state. The transformed state results in a superposition of two-mode photon-subtracted Fock states, and therefore it displays non-Gaussian statistics, more specific a non-Gaussian probability distribution in quadrature space. In particular, the transformed state consist of a quantum vortex structure, meaning a quantum state whose wave-function takes the general form \cite{34}:
\begin{equation}
\psi(x,y)=(x-iy)^{m} e^{(x^2-y^2)/2\sigma^2},
\end{equation}
where $m$ is an integer. It is well known that quantum vortex states are a particular class of non-Gaussian quantum states \cite{34,34b}, and are therefore amenable to applications requiring to venture into the non-Gaussian domain. Moreover, we show that the non-Gaussian quantum state we propose presents enhanced entanglement content, as quantified by the enhanced Logarithmic Negativity of their Wigner function probability distribution. We argue that such non-Gaussian quantum states can find relevant applications in the context of quantum information protocols requiring non-Gaussian resources, in particular in  entanglement distillation schemes. \\

The article is structured as follows: In Section II, we introduce our technique for deterministic preparation of non-Gaussian quantum vortex states on-demand. Second, in Section III, we present an exhaustive investigation of the Wigner function representation of the resulting non-Gaussian quantum vortex state. In Section IV, we present a complete study of the entanglement content in the proposed non-Gaussian quantum states, demonstrating that the non-Gaussian nature of the quantum vortex state enhances the quantum correlations existing in the initial two-mode squeezed Gaussian state, and are therefore  suitable candidates for quantum information protocols. Next, in Section V, we propose a concrete application of such non-Gaussian quantum states in quantum distillation protocols. Finally, in Section VI we outline our conclusions. \\

\section{Non-Gaussian Quantum Vortex states}

\begin{figure}[t!]
\includegraphics[width=0.5\textwidth]{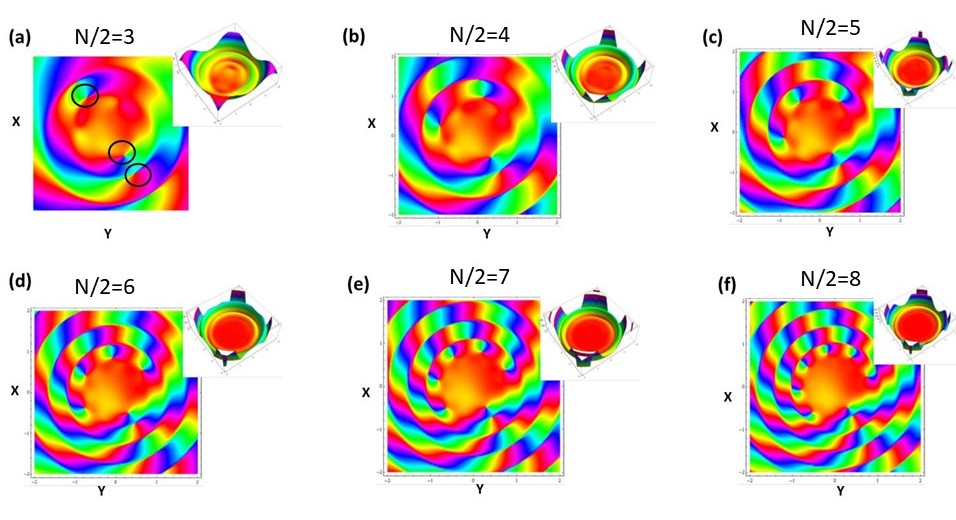}
\caption{Phase profile of resulting non-Gaussian quantum vortex state in quadrature representation for a squeezing parameter r = 0.02, exploring the impact of the photon-number ($N/2$ per mode) in the formation of vortices. Insets correspond to amplitude plots  The numerical results confirm creation of $N/2$ vortices for N total input photons. (a) $N/2 = 3$, (b) $N/2 = 4$, (c) $N/2 = 5$, (d) $N/2 = 6$, (e) $N/2 = 7$, (f) $N/2 = 8$, with $N/2$ input photons per mode (figure reproduced under the terms and conditions of the Creative Commons Attribution License \cite{35}).}
\label{Fig:quadrature distribution}
\end{figure} 

The non-Gaussian quantum vortex states that we deal with in this article are generated from truncated two-mode photon-number squeezed states. In the Fock representation, these states are written as

\begin{equation}
    \label{Eq:Photon-number squeezed state}
    \vert \psi \rangle = \frac{A}{\cosh{r}}\sum_{j=0}^N\left(\tanh{r}\right)^j \vert j \rangle_a \vert j \rangle_b
\end{equation}

\noindent where $(a,b)$ are mode labels, $r$ is the squeezing parameter and $A$ is an additional normalization factor. Similar states are routinely generated in spontaneous parametric down conversion (SPDC) processes where generally an infinite series is considered with diminishing probability of the higher order photon-number terms. In fact, the pump power in such processes is kept sufficiently small to ensure significantly less chances of occurrence of the higher order terms. In such scenario, Eq. 2 
can be treated as the more practical representation of SPDC output. The additional normalization factor is significant in this case since it arises due to such truncation of the infinite series. In what follows, we treat $(a,b)$ as distinct spatial modes.\\

\par Next, we want to study the effect of beam splitter transformation on Eq. 2. The two spatial modes are directed to the two different inputs of the beam splitter. We map the input modes $(a,b)$ to the output modes $(a',b')$ of the beam splitter as follows
\begin{eqnarray}
\label{Eq:Beam splitter transformations}
a &\rightarrow & a' = \frac{1}{\sqrt{2}}\left(a - ib\right) \\
b &\rightarrow & b' = \frac{1}{\sqrt{2}}\left(b - ia\right)
\end{eqnarray}

This transforms the initial photon-number squeezed state to the following

\begin{eqnarray}
    \label{Eq:Quantum vortex state}
    \vert \psi'\rangle &=& \frac{A}{\cosh{r}}\sum_{j=0}^N \tanh{r}^j\frac{j!}{2^{N/2}}\nonumber \\
   &\times & \sum_{k,l}^j i^{l+k}C_{k,l}\vert j - (l-k)\rangle \vert j + (l-k)\rangle
\end{eqnarray}

\noindent where $C_{k,l} = \frac{\sqrt{(j-l+k)!(j+l-k)!}}{k!(j-k)!l!(j-l)!}$. We call $\vert \psi'\rangle$ the non-Gaussian quantum vortex state. The transformed state can be regarded as a superposition of photon-subtracted Fock states, due to the imbalance in photon-number between the two modes, therefore its non-Gaussian character is apparent. Since a beam splitter transformation is a unitary process, there is no change in the total number of photons in the two modes combined between the photon-number squeezed state $\vert \psi\rangle$ and the non-Gaussian quantum vortex state. Rather, it results in a redistribution of the photons between the two modes for each term of the summation in Eq. \ref{Eq:Photon-number squeezed state}. The nomenclature would become clear if we look at the associated quadrature distribution. Using the relation
\begin{equation}
    \label{Eq:Quadrature 1}
    \psi_n\left(x\right) = \langle x \vert n \rangle = \frac{1}{\sqrt{2^n n!\sqrt{\pi}}}\exp\left(-\frac{x^2}{2}\right)H_n(x)
\end{equation}

\noindent where $H_n(x)$ is the Hermite polynomial of order $n$ and $\vert n \rangle$ is a number state, and noting that product of Hermite polynomials $H_n(x)H_m(y)$ can be written in terms of Laguerre polynomials $L^{n-m}_m(x^2+y^2)$, the quadrature distribution can be derived as follows \cite{35}

\begin{eqnarray}
    \label{Eq:Quadrature distribution}
    \psi'(x,y) &=& \frac{A'}{\cosh{r}}\sum_{j=0}^N\sum_{m=0}^j\tanh{r}^j\frac{j!}{2^{N/2}}C_{k,l}\nonumber \\
    &\times & L_{j-m}^{2m}\left(x^2+y^2\right)\exp\left(-\frac{x^2+y^2}{2}\right)
\end{eqnarray}

\noindent where $m=l-k$. This illustrates the fact the state after the beam splitter transformation consists of a finite of quantum vortices of order $2m$. This justifies the nomenclature. As a further illustration, we showcase a few cases of quantum vortex states in Fig. \ref{Fig:quadrature distribution}. It is quite evident from the amplitude plots that these states exhibit a departure from their initial Gaussian-like characteristics. 

\section{Wigner Function Representation of Non-Gaussian Quantum Vortex States}

In the previous section, we have outlined the method to generate the non-Gaussian quantum vortex states and looked at their associated quadrature distribution. In this section, we want to study their phase-space distribution to better understand the non-Gaussian structure inherent in these states. In order to do that, we will use the Wigner distribution function. The Wigner distribution is a quasi-probability distribution function that is real, non-singular and produces accurate quantum mechanical operator averages. It is also characterised by a definite marginal distribution. A remarkable advantage of using the Wigner function is that it can reveal both non-classicality and non-Gaussianity of quantum states. 
\begin{figure*}
\centering
\includegraphics[width=0.8\textwidth]{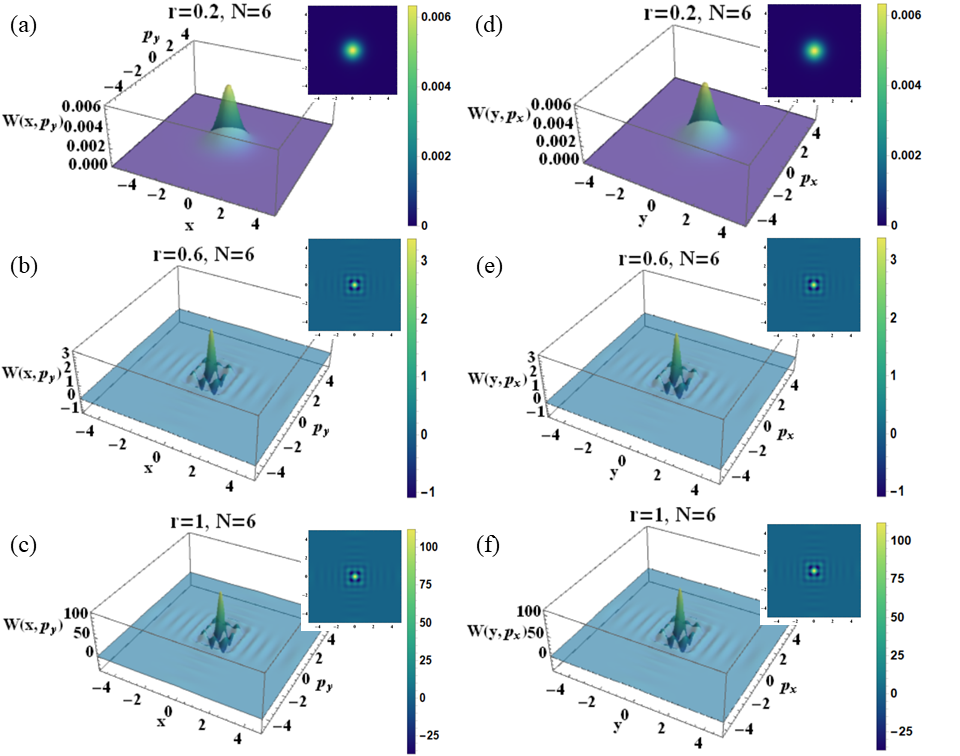}
\caption{Two dimensional slice of the Wigner function of the non-Gaussian quantum vortex state for $N = 6$ and different values of $r$. The non-Gaussian nature is evident from the negative regions in the above plots. It is to be noted that the negativity arises and then increases with increasing $r$. (a), (b) and (c) are projections of the Wigner function at the plane $\left(y=0,p_x=0\right)$ while (d), (e) and (f) are projections of the Wigner function at the plane $\left(x=0,p_y=0\right)$. Inset are the corresponding contour plots for each of the cases. Details are in the text.}
\label{Fig:Wigner_panel_1}
\end{figure*}
\par For a single mode quantum state $\vert n \rangle$ in the Fock space representation, the associated Wigner function is defined as

\begin{equation}
    \label{Eq:Wigner}
    W\left(x,p_x\right) = \frac{2}{\pi}\left(-1\right)^nL_n(4q^2)e^{-2q^2}
\end{equation}

\noindent where $q^2=x^2 + p_x^2$ and $L_n(4q^2)$ is again Laguerre polynomial of order $n$. $\left(x,p_x\right)$ are the quadrature variables. The Wigner function of the two-mode state $\vert n \rangle_a\vert m \rangle_b$ is then simply the product of the corresponding Wigner functions as follows

\begin{eqnarray}
    \label{Eq:Wigner 2 mode}
    W\left(x,p_x,y,p_y\right) &=& \frac{4}{\pi^2}\left(-1\right)^{n+m}\nonumber \\
    &\times & L_n(4q_a^2)e^{-2q_a^2}L_m(4q_b^2)e^{-2q_b^2}
\end{eqnarray}

\noindent where $q_a^2=x^2 + p_x^2$, $q_b^2=y^2 + p_y^2$ while $\left(x,p_x\right)$ and $\left(y,p_y\right)$ are the quadrature variables for modes $a$ and $b$ respectively.
\begin{figure*}
\includegraphics[width=0.8\textwidth]{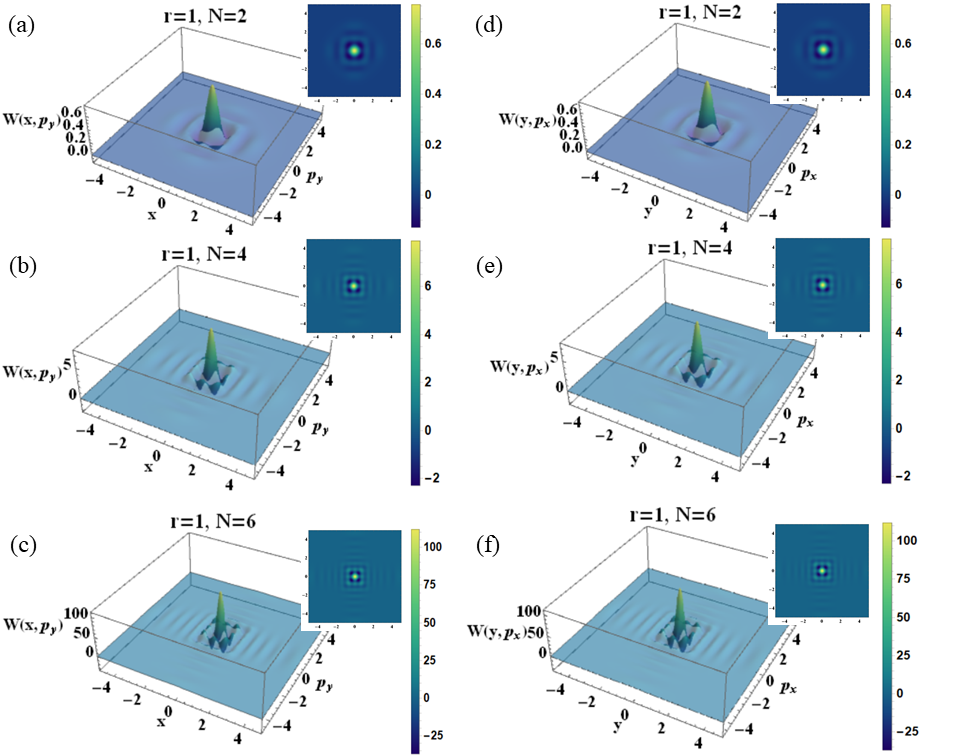}
\caption{Two dimensional slice of the Wigner function of the non-Gaussian quantum vortex state for a fixed value of $r$ and different value of $N$. The non-Gaussian nature is again evident from the negative regions in the above plots. (a), (b) and (c) are projections of the Wigner function at the plane $\left(y=0,p_x=0\right)$ while (d), (e) and (f) are the projections of the Wigner function at the plane $\left(x=0,p_y=0\right)$. Inset are the corresponding contour plots. Details are in the text.}
\label{Fig:Wigner_panel_2}
\end{figure*}
\par In our case, the non-Gaussian quantum vortex state is obtained by performing an unitary operation on an initial state in photon-number superposition. Using its Fock space representation of Eq. (\ref{Eq:Quantum vortex state}), after some manipulation, we can write the Wigner function in the following form

\begin{equation}
    \label{Eq:Final Wigner}
    W=\frac{A''}{\cosh{r}^2}\sum_{j=0}^N\sum_{m=0}^j\tanh{r}^{2*j}\frac{j!^2}{2^{N}}\vert C_{k,l}\vert ^2 W_{j,m}
\end{equation}

\noindent where $A''$ is the normalization term and $W_{j,m}$ are defined as

\begin{eqnarray}
    \label{Eq:Wigner LG}
    W_{j,m} &=& \frac{(-1)^{2*j}}{\pi}L_{j+m}[4(Q_0+Q_1)] \nonumber \\
    &\times & L_{j-m}[4(Q_0-Q_1)]\exp{\left(-4Q_0\right)}
\end{eqnarray}

\noindent where $Q_0=\frac{1}{4}\left(x^2+y^2+p_x^2+p_y^2\right)$ and $Q_1=\frac{xp_y - yp_x}{2}$. We study projections of the Wigner function of the non-Gaussian quantum vortex state on different planes in Fig. \ref{Fig:Wigner_panel_1} and Fig. \ref{Fig:Wigner_panel_2}. Since the Wigner function is 4-dimensional for a two-mode state, it is impossible to graphically reproduce the entire structure.We therefore chose to highlight only those planes which best illustrates the negative volume of the Wigner function. As can be seen from the figures, the projections at the planes $\left(y=0, p_x=0\right)$ and $\left(x=0, p_y=0\right)$ exhibit maximum negativity. The negative volume increases with increasing $r$ and $N$ though, the squeezing parameter has greater influence. Even for high values of $N$, there is no negative region for low values of $r$ as is shown in Fig, 2a and 2d. Comparing this with Fig. 3a and 3d, we see that even for the case with minimum number of photons, negative regions are present for sufficiently high values of $r$. In order to better quantify these results, we compute the total negative volume of the entire Wigner function below.
\par
\begin{figure}[b!]
\includegraphics[width=0.45\textwidth]{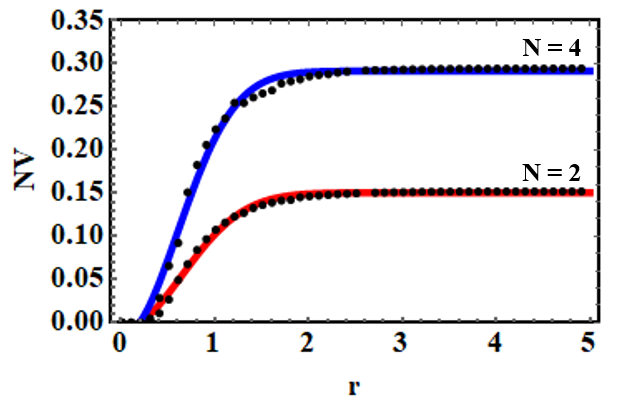}
\caption{Negativity volume of the Wigner function of the non-Gaussian quantum vortex states for different values of $N$. The black markers are the datapoints from numerical evaluation of Eq. \ref{Eq:Wigner negativity}. The continuous curves are numerical best fits.}
\label{Fig:Wigner negativity}
\end{figure}

 Now, as has been mentioned before, an important aspect of the Wigner function is its usefulness in detecting nonclassicality and non-Gaussianity of quantum states. In general, the presence of negative regions in the Wigner function is accepted as a signature of nonclassicality. Additionally, the Wigner function for a pure quantum state that is Gaussian is strictly positive \cite{HudsonRMP}. This means that the negativity of the Wigner function is a witness of non-Gaussianity. This can be quantified with the help of the negativity volume of the Wigner function defined as

\begin{equation}
    \label{Eq:Wigner negativity}
    \mathrm{NV}=\frac{1}{2}\left(\int_V |W| \text{d}V - 1\right)
\end{equation}

\noindent where the integration is performed over the entire phase space. We have performed a numerical integration to calculate the negative volume of the non-Gaussian quantum vortex states. We study the results in Fig. \ref{Fig:Wigner negativity}.
\par We see that the negative volume increases with increasing values of $r$ and saturates to a maximum value. This has two related interpretations. It means that increased squeezing leads to increases non-classicality which is already known. In addition, it also signifies that an increase in the squeezing parameter also leads to increased non-Gaussianity of the final state. 
\begin{figure*}
\includegraphics[width=\textwidth]{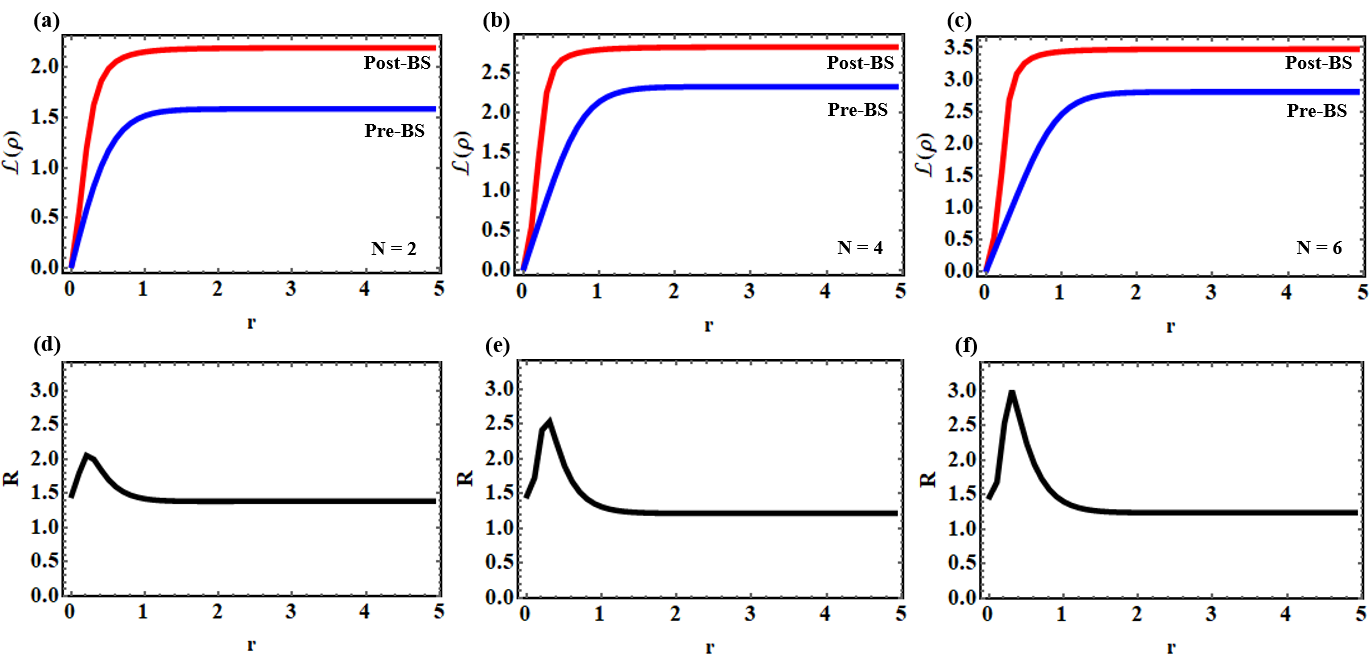}
\caption{\emph{Top row}: Logarithmic negativity of the non-Gaussian quantum vortex states and two-mode photon-number squeezed states for different values of N in arbitrary units. \emph{Bottom row}: Ratio of the logarithmic negativities for the two states better illustrates the increase in entanglement due to the beam splitter transformation. Here $R = \frac{\mathcal{L}\left(\rho\right)}{\mathcal{L}\left(\rho'\right)}$ where $\rho'$ is the density matrix of the two-mode photon-number squeezed state before the beam splitter transformation. (a) and (d) $N = 2$, (b) and (e) $N = 4$, (c) and (f) $N = 6$.}
\label{Fig:LogNegativity}
\end{figure*}
We should also point out that the negativity increases with increasing number of photons as is evident from the plots (red for $N=2$ and blue for $N=4$). This means that the redistribution of photon number between the two modes effected by the beam splitter transformation is successful in increasing the nonclassical effects of the photon-number squeezed states. More interestingly, it shows that the transformation also leads to the generation of significant non-Gaussianity.
\par The above results serve as a pointer to investigate how the entanglement between the two modes are influenced by this redistribution of photon numbers which we study in the next section.

\section{Entanglement Content of Non-Gaussian Quantum Vortex States}

In this section we study the entanglement content of non-Gaussian quantum vortex states. We use Logarithmic Negativity \cite{Plenio2005} which is easily computable and a proven entanglement monotone. It is defined as

\begin{equation}
    \label{Eq:Log Negativity}
    \mathcal{L}\left(\rho\right) = \log_2 \left(1 + 2 \mathrm{N}\right)
\end{equation}

\noindent where $\rho$ is the density matrix of the quantum state and $\mathrm{N}$ is the negativity measure \cite{Vidal2002}, defined as

\begin{equation}
    \label{Eq:Negativity}
    \mathrm{N} = \frac{||\rho^T || - 1}{2}
\end{equation}

\noindent Here $||.||$ denotes trace norm and $\rho^T$ is the partial transpose of the density matrix of the quantum state with respect to one of the subsystems. The negativity then is the absolute values of the sum of all negative eigenvalues of the partially transposed density matrix. In order to calculate $\mathcal{L\left(\rho\right)}$ of the non-Gaussian quantum vortex state, we first need to write the density matrix corresponding to $\vert \psi' \rangle$. It has the form

\begin{eqnarray}
    \label{Eq:Density matrix}
    \rho &=& \vert \psi'\rangle\langle \psi' \vert \nonumber \\
    &=& \frac{|A|^2}{\cosh^2 r}\sum_{j,j'}^N\tanh{r}^{j+j'}\frac{j!j'!}{2^N}\sum_{k,l,k',l'}^{j,j'}i^{k+l}(-i)^{k'+l'}C_{k,l}C_{k',l'}\nonumber \\
    &\times & \vert j-(l-k)\rangle\langle j'-(l'-k')\vert\otimes \vert j+(l-k)\rangle\langle j'+(l'-k')\vert \nonumber \\
    ~
\end{eqnarray}

The partial transpose of Eq. \ref{Eq:Density matrix} is obtained by standard techniques and then diagonalized numerically to calculate the negative eigenvalues which are then used to determine $\mathcal{L}(\rho)$. A similar process is followed to calculate the logarithmic negativity of the two-mode photon-number squeezed state of Eq. 2 for which the density matrix has the form

\begin{equation}
    \label{Eq:Density matrix 1}
    \rho' =  \frac{|A|^2}{\cosh^2 r}\sum_{j,j'}^N\left(\tanh{r}\right)^{j+j'}\vert j\rangle\langle j'\vert \otimes \vert j\rangle\langle j'\vert
\end{equation}

We compare the values of $\mathcal{L}\left(\rho\right)$ and $\mathcal{L}\left(\rho'\right)$ as well as study their ratio $R$ as a function of squeezing parameter $r$ in Fig. \ref{Fig:LogNegativity}.
\par We see that the entanglement increases after the beam-splitter operation. This is due to the fact that the uncertainty in photon number in each mode increases due to the redistribution of photons between the two modes mediated by the beam-splitter. The total entanglement increases with increasing number of photons as well. Also, as $r$ increases, the entanglement increases sharply before saturating to a maximum value. More importantly, entanglement vanishes in the absence of squeezing since there are no photons present in either modes in such situation.

\section{Applications of Non-Gaussian Quantum Vortex States in Quantum Information Protocols}

\subsection{Entanglement Distillation Protocols}

For finite-dimensional systems, the term entanglement distillation has been linked to the notion that one can obtain highly entangled states by means of local quantum operations and classical communication, by starting from a large number of weakly entangled quantum states and ending with a smaller number of more entangled ones. Such approaches also work as the basis for quantum cryptographic schemes. An equivalent procedure should also exists for the distillation of Gaussian states by means of local Gaussian operations and classical communication \cite{37}. At any rate, entanglement distillation aims at producing more highly entangled states out of a situation where entanglement is present only in a noisy form,
presumably as a consequence of some lossy quantum channel. Entanglement distillation can be regarded as a key element in quantum
repeater approaches, allowing for  long-range entanglement distribution in the presence of noise.
In essence, it is possible to differentiate between  distillation protocols that involve several copies of an entangled state at each step of the scheme, and local
filtering approaches that take a single specimen of a state and, under appropriate filtering, 
give rise to a more entangled state. In the context of Gaussian
operations, CV entanglement distillation of neither kind is possible without the aid of non-Gaussian operations, such as photon addition
or subtraction \cite{37}.

\subsection{Proposed Experimental Scheme}

\begin{figure}[b!]
\includegraphics[width=0.5\textwidth]{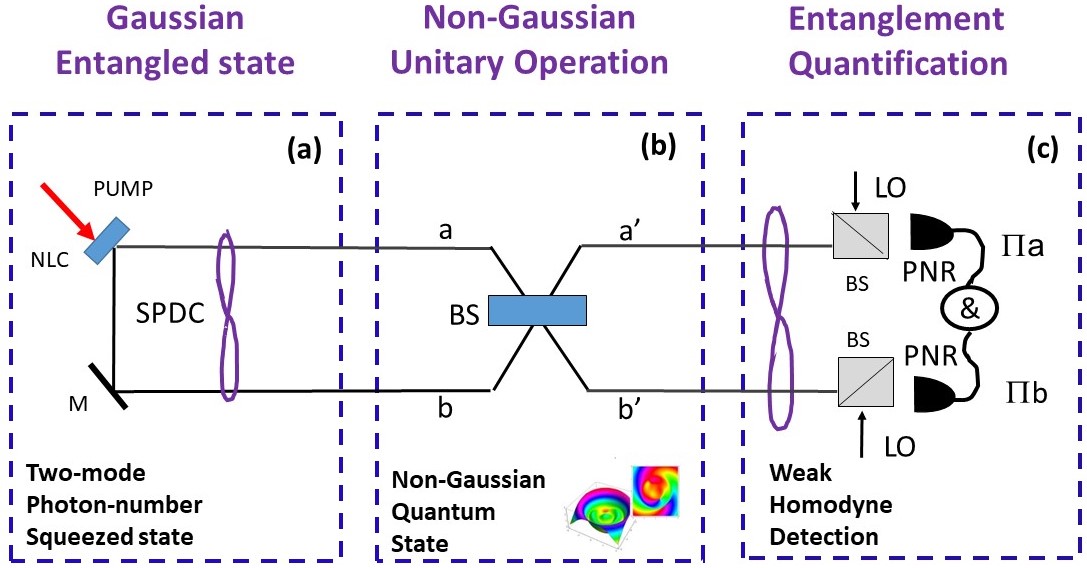}
\caption{Proposed experimental scheme for implementation of entanglement distillation protocol. (a) two-mode Gaussian entangled state  produced in a non-linear crystal (NLC) via non-collinear SPDC. (b) Deterministic non-Gaussian quantum state prepared via transmission through a beam-splitter BS. ($a,b$) and ($a',b'$) label the input and output spatial modes transmitted by the BS, respectively. Such unitary operation enhances the entanglement content. (c)  Entanglement distillation 
quantified by a
partial detection approach via weak homodyne detection using photon-number-resolving (PNR) detectors and weak local oscillators (LO). Selected joint POVM elements $\Pi_{ab}$ can provide for accurate bounds on the entanglement content via Convex Optimization approaches \cite{36} (see text for details).}
\end{figure}

In order to distill CV entanglement from Gaussian states, such as the two-mode photon-number squeezed state described by in Section II, an operation that removes the
Gaussianity of the field quadrature statistics is required. Examples of such non-Gaussian
operations include the conditional subtraction or addition of a photon [1-11]. However, one main limitation of such conditional operations is their probabilistic nature. Here we propose a deterministic approach for removing the Gaussian statistics of the initial two-mode photon-number squeezed state, such deterministic unitary operation consists of transmission through a standard beam-splitter (BS). It has recently been demonstrated that such operation introduces photon-number fluctuations \cite{35}, which result in the creation of quantum vortices in the quadratures with non-Gaussian statistics, therefore creating a non-Gaussian quantum state suitable for applications in quantum information protocols, such as entanglement distillation. \\

The proposed experimental scheme for entanglement distillation is depicted in Fig. 6. It consists of three main steps. The first step (Fig. 6 (a)) is the preparation of the initial Gaussian entangled state, via non-collinear SPDC, using a non-linear crystal (NLC). As anticipated in Section II, for a sufficiently attenuated pump the resulting state can be approximated by a truncated two-mode photon-number squeezed Gaussian (G) quantum state of the form \cite{36}:

\begin{equation}
    \vert \psi \rangle_{\mathrm{G}} = \frac{A}{\cosh{r}}\sum_{j=0}^N\left(\tanh{r}\right)^j \vert j \rangle_a \vert j \rangle_b
\end{equation}

where $A$ is a normalization constant and  $r$ the squeezing parameter.\\  

The second step (Fig. 6 (b)) consists of a non-Gaussian unitary operation, meaning an operation that removes the Gaussian measurement statistic of field quadratures in the initial state, resulting in a non-Gaussian (NG) quantum state. As described in detail in Ref. \cite{35}, the beam-splitter introduces photon-number fluctuations resulting in a quantum state with a binomial photon-number distribution of the form:
\begin{eqnarray}
|\psi \rangle_{\mathrm{NG}} & = & \frac{D}{\cosh {r}} \sum_{j=0} ^{N/2} A^{r,N}_{j} \times \\
& & \sum_{k=0} ^{j} \sum_{l=0} ^{j} B^{\phi}_{k,l} C^{N j}_{k l} |j-(l-k), j+(l-k) \rangle. \notag
\end{eqnarray}
where $D$ is the normalization factor. Explicit expressions for the coefficients $A^{r,N}_{j}, B^{\phi}_{k,l}, C^{N j}_{k l} $ are given in Section II. Such states exhibits a vortex structure in quadrature space (Fig. 1), and non-Gaussian field quadrature statistics.  \\

As reported in Section III and IV, the resulting  non-Gaussian state  displays enhanced entanglement content, which is later on quantified by a partial detection approach, based on Entanglement Witnesses and Convex Optimization schemes  \cite{36}. The non-Gaussian quantum state is eventually routed towards a weak homodyne detection station, using standard silver mirrors ($M$).\\

The final step (Fig.6 (c)) in the entanglement distillation protocol consists of a partial detection approach, implemented by constructing suitable Entanglement Witnesses, which in turn are built by selecting suitable joint Positive Operator Valued Measures (POVM) $\Pi_{ab}=\Pi_{a} \otimes \Pi_{b}$ for each detector $D_{a,b}$, where $a,b$ label each spatial mode transmitted by the BS. Photons are detected using photon number resolving (PNR) detectors and weak local oscillators. The POVM elements of such weak homodyne detectors have been fully characterized in Ref \cite{38}. The measurement outcomes provided by the selected POVM elements can provide for accurate bounds on the entanglement content via Convex Optimization approaches. A full description of such Convex Optimization schemes is reported in Ref. \cite{36}.

\section{Discussion}

We presented a scheme for deterministic preparation of non-Gaussian quantum states on-demand. In contrast to the standard  probabilistic approaches for preparation of non-Gaussian states, conditioned on photon subtraction or addition, the scheme presented here can  prepare non-Gaussian quantum states on-demand by applying a deterministic unitary transformation which removes the Gaussianity of quadrature statistics of the initial state, namely a quadrature rotation via transmission through a standard beam-splitter using a  photon-number squeezed state as input. The resulting non-Gaussian quantum state consists of a superposition of photon-subtracted Fock states, and it exhibits a quantum vortex structure in quadrature space, thus confirming the non-Gaussian character of measurement statistics of field quadratures, it also reveals increased entanglement content, as quantified by the Logarithmic Negativity and the Wigner function negative volume, therefore displaying high potential for applications in quantum information protocols such as  entanglement distillation schemes. 

\section{Acknowledgements}
The authors gratefully acknowledge Jens Eisert for helpful discussions. G.P. acknowledges financial support via PICT Startup.

\end{document}